\documentclass{LT23auth}
\usepackage[dvips]{graphicx}

\begin{document}

\begin{frontmatter}

\title{Neutral Collective Excitations in Striped Hall States}

\author{N. Maeda\thanksref{thank1}},
\author{T. Aoyama},
\author{Y. Ishizuka},
\author{K. Ishikawa}

\address{Department of
Physics, Hokkaido University, Sapporo 060-0810, Japan.}

\thanks[thank1]{Corresponding author. 
E-mail: maeda@particle.sci.hokudai.ac.jp}

\begin{abstract}
In the striped Hall state, a magnetic translation in one direction is 
spontaneously broken to the discrete translation. 
The spectrum of the neutral collective excitation is obtained in 
the single mode approximation at half-filled third and 
fourth Landau levels. 
The spectrum is anisotropic and has a multiple line node structure. 
In one direction, the spectrum resembles the liquid Helium 
spectrum with the phonon and roton minimum. 
\end{abstract}

\end{frontmatter}


Recently, highly anisotropic states were observed around the 
half-filled third and higher Landau levels\cite{Lilly}.  
The anisotropic state is believed to be the striped Hall state which 
is a unidirectional charge density wave in a mean field theory\cite{Kou,Moe}.  
The anisotropy is naturally explained by the 
anisotropic Fermi surface in the magnetic Brillouin zone\cite{Imo}. 
It is predicted that the fluctuation effect turns the striped state 
into the smectic or nematic liquid crystal\cite{Fra} using 
the Hartree-Fock approximation or coupled Luttinger liquids theory. 

In this paper, we investigate the property of the neutral collective 
excitations in the striped Hall state. 
In the absence of edges and disorder, 
a two-dimensional electron system under a uniform magnetic field 
has the magnetic translation and rotation symmetry. 
In the striped Hall state, a magnetic translation in one direction is 
spontaneously broken to the discrete translation and the rotation is also 
spontaneously broken to the $\pi$-rotation. 
Goldstone theorem for the striped Hall state\cite{sma} states 
that the gapless excitation exists in the 
neutral charge sector and couples with the density operator. 
The spectrum of the neutral collective excitation is obtained in 
the single mode approximation (SMA) numerically. 
We use the unit $\hbar=c=1$ and $a=\sqrt{2\pi\hbar/eB}=1$. 

In a strong magnetic field $B$, the free kinetic energy is quenched. 
Therefore we study only the interaction Hamiltonian projected into 
the l th Landau level, 
\begin{eqnarray}
H_l&=&{1\over2}\int d{\bf r}d{\bf r}'\Psi^\dagger({\bf r})
\Psi^\dagger({\bf r}')V({\bf r}-{\bf r}')\Psi({\bf r}')\Psi({\bf r}),
\nonumber\\
\Psi({\bf r})&=&\int_{\rm BZ}{d^2p\over(2\pi)^2}
b_l({\bf p})\langle {\bf r}\vert l,{\bf p}\rangle,
\end{eqnarray}
where $\Psi$ is the electron field operator projected into 
the l th Landau level\cite{Imo}, BZ stands for Brillouin zone 
$\vert p_i\vert<\pi$, and $V({\bf r})=q^2/r$ 
($q^2=e^2/4 \pi \epsilon $, $\epsilon$ is the dielectric constant).
$b_l({\bf p})$ is an annihilation operotor for one-particle state 
$\langle {\bf r}\vert l,{\bf p}\rangle$ which is a Bloch wave on 
the magnetic von Neumann lattice\cite{Imo}. 

The mean field state for the striped Hall state is constructed as
\begin{equation}
\vert{\rm stripe}\rangle=\prod_{{\bf p}\in {\rm F.S.}}
b_l^\dagger({\bf p})\vert 0\rangle,
\label{mfs}
\end{equation}
where F.S. means Fermi sea $\vert p_y\vert<\pi/2$, and 
$\vert 0\rangle$ is the state in which the $l-1$ th and 
lower Landau levels are fully occupied. 
The charge density for state of Eq.~(\ref{mfs}) is uniform 
in $y$ direction and periodic in $x$ direction. 
The period $r_s$ is a parameter of the von Neumann lattice 
and is fixed by the minimum energy condition\cite{Imo}. 

We calculate the spectrum for a neutral collective 
excitation at the half-filled third and fourth Landau level 
using SMA, which was successful 
in the FQHS because 
the backflow problem is absent for the electron states projected 
to the Landau level\cite{Gir}. 
Projected density operator $\rho({\bf k})$ is written as 
$e^{-k^2/8\pi}L_l(k^2/4\pi)\rho_*({\bf k})$, where 
\begin{equation}
\rho_*({\bf k})=\int_{\rm BZ}{d^2p\over(2\pi)^2}b_l^\dagger({\bf p})
b_{l}({\bf p}-\hat{\bf k})e^{-{i\over4\pi}{\hat k}_x(2p_y-{\hat k}_y)},
\end{equation}
where $\hat{\bf k}=(r_s k_x,k_y/r_s)$. 
It is well-known that the density operators projected to the 
Landau level satisfies, 
\begin{equation}
[\rho_*({\bf k}),\rho_*({\bf k}')]=-2i\sin\left(
{{\bf k}\times{\bf k}'/4\pi}\right)
\rho_*({\bf k}+{\bf k}').
\label{rho}
\end{equation}

The variational excited state is defined by $\vert{\bf k}\rangle=
\rho_*({\bf k})\vert{\rm stripe}\rangle$ and 
the variational excitation energy $\Delta({\bf k})$ is written as
\begin{eqnarray}
\Delta({\bf k})&=&{\langle{\bf k}\vert (H_l-E_0)\vert{\bf k}\rangle
\over\langle{\bf k}\vert{\bf k}\rangle}=
{f({\bf k})\over s({\bf k})},\nonumber\\
f({\bf k})&=&\langle0\vert [\rho_*(-{\bf k}),[H_l,\rho_*({\bf k})]]
\vert0\rangle/2N_e^*,\\
s({\bf k})&=&\langle0\vert\rho_*(-{\bf k})\rho_*({\bf k})
\vert0\rangle/N_e^*,\nonumber
\end{eqnarray}
where $E_0$ is a ground state energy, $N_e^*$ is a electron number 
in the l th Landau level, and 
$s({\bf k})$ is the so-called static structure factor. 
To derive these expressions, we use the relation $f(-{\bf k})=f({\bf k})$ 
and $s(-{\bf k})=s({\bf k})$ due to $\pi$ rotation symmetry. 
Using the commutation relation (\ref{rho}), $f({\bf k})$ is written as 
\begin{eqnarray}
f({\bf k})&=&2\int{d^2k'\over(2\pi)^2}v_l(k')\sin^2
\left({{\bf k}'\times{\bf k}\over4\pi}\right)\\
&&\times\{s({\bf k}+{\bf k}')-s({\bf k}')\},\nonumber
\end{eqnarray}
where $v_l(k)=e^{-k^2/4\pi}(L_l(k^2/4\pi))^2 2\pi q^2/k$. 
Therefore the variational excitation energy is calculable if we know 
the static structure factor $s({\bf k})$. 
For the mean field state (\ref{mfs}), 
$s({\bf k})$ behaves as $\vert k_y\vert/\pi r_s$ at small 
$k_y$ and periodic in $k_y$ direction with a period $2 \pi r_s$. 
The numerical results for the energy spectrum $\Delta$ for $\nu=l+1/2$, 
$l=2$ and $3$ are shown in Figs.~1 and 2, respectively. 

\begin{figure}[btp]
\begin{center}\leavevmode
\includegraphics[width=0.9\linewidth]{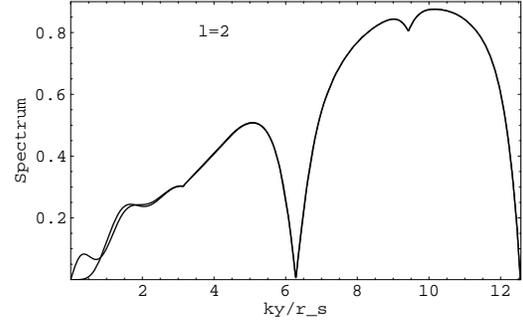}
\caption{ 
Energy spectrum $\Delta$ at $0<r_s k_y<4\pi$ 
for $k_x=0$ and 1 (linear dispersion at $k_y=0$), 
$\nu=2+1/2$ in SMA. 
The unit of $\bf k$ is $a^{-1}$ and the unit of spectrum is 
$q^2/a$. The same unit is used in Fig.~2. 
}\label{figurename1}\end{center}
\end{figure}
\begin{figure}[btp]
\begin{center}\leavevmode
\includegraphics[width=0.9\linewidth]{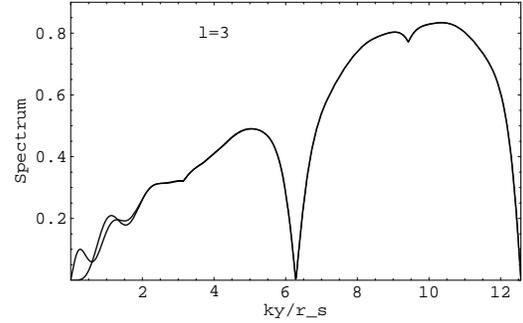}
\caption{ 
Energy spectrum $\Delta$ at $0<r_s k_y<4\pi$ 
for $k_x=0$ and 1 (linear dispersion at $k_y=0$), 
$\nu=3+1/2$ in SMA. 
}\label{figurename2}\end{center}
\end{figure}

As seen in these figures, the spectrum in $k_y$ direction 
resembles the liquid Helium spectrum with the phonon and roton minimum. 
The spectrum has a multiple line node at $k_y=2\pi r_s n$ 
($n$ is integer). 
In a usual Fermi system, SMA does 
not work well because of the particle-hole excitation near the 
Fermi surface. 
In the present case, however, the particle-hole excitation is 
suppressed by the divergence of the Fermi velocity due to 
the Coulomb interaction. 
Actually, the comparison to particle-hole excitation energy\cite{sma} 
shows that SMA is good around 
$k_y=2\pi r_s n$. 
We hope these spectrum will be observed for the evidence 
of the striped Hall state. 

This work was partially supported by the special Grant-in-Aid for 
Promotion of Education and Science in Hokkaido University, and by 
the Grant-in-Aid for Scientific Research on Priority area of Research (B) 
(Dynamics of Superstrings and Field Theories, Grant No. 13135201),  
provided by the Ministry of Education, Science, Sport, and Culture, Japan. 

%
%

\end{document}